\documentclass[prl,twocolumn,showpacs,groupedaddress]{revtex4}

\usepackage{graphicx}

\begin{document}

\title{\bf Direct Evidence for a Two-component Tunnelling Mechanism \\
in the Multicomponent Glasses at Low Temperatures }
 
\author{ Giancarlo Jug$^1$ and Maksym Paliienko }

\affiliation{ 
Dipartimento di Fisica e Matematica, Universit\`a dell'Insubria, Via
Valleggio 11, 22100 Como, Italy \\
$^1$CNISM -- Unit\`a di Ricerca di Como and INFN -- Sezione di Pavia, 
Italy }   

\date{\today }

\begin{abstract}
The dielectric anomalies of window-type glasses at low temperatures 
($T<$ 1 K) are rather successfully explained by the two-level systems
(2LS) tunneling model (TM). However, the magnetic effects discovered in 
the multisilicate glasses in recent times \cite{ref1}-\cite{ref3}, and 
also some older data from mixed (SiO$_2$)$_{1-x}$(K$_2$O)$_x$ and 
(SiO$_2$)$_{1-x}$(Na$_2$O)$_x$ glasses \cite{ref4}, indicate the need 
for a suitable generalization of the 2LS TM. We show that, not only for 
the magnetic effects \cite{ref3,ref5} but also for the mixed glasses in 
the absence of a field, the right extension of the 2LS TM is provided by 
the (anomalous) multilevel tunneling systems approach proposed by one of 
us. It appears that new 2LS develop via dilution near the hull of the 
SiO$_4$-percolating clusters in the mixed glasses.   
\end{abstract}


\maketitle

Tunneling systems are useful for many theoretical investigations in
condensed-matter physics: from laser-matter interaction, to amorphous
solids and disordered crystals, to hydrogen storage in metals. For their 
simplicity, only the two-level systems (2LS) are usually employed instead 
of the more realistic multilevel systems; in this Letter a situation will 
be discussed where the use of a number of states greater than two is
essential. Moreover, new insight will be given on the role of percolation 
and fractal theory in the TM of multicomponent glasses.  

Glasses at low temperatures are interesting for their rather universal 
physical properties attributed to the low-energy excitations characterising 
all kinds of amorphous solids. The 2LS tunneling model (TM) has been 
successful in the semi-quantitative explanation of a variety of interesting 
thermal, dielectric and acoustic anomalies of structural glasses at 
temperatures $T<$ 1 K \cite{ref6}. The physics of cold glasses is important
also because of its link with the mechanism of the glass freezing transition 
\cite{ref7}. Beside the linear in $T$ behavior of the heat capacity $C_p$, 
it is believed that the linear in $\pm\ln T$ behavior of the real-part of 
the frequency-dependent dielectric constant $\epsilon'(\omega)$ represents a 
cogent characterization of the glassy state at low temperatures.   

Fig. 1 (inset) shows the behavior of the $T$-dependent part of $\epsilon'$, 
$\Delta\epsilon'/\epsilon'=[\epsilon'(T)-\epsilon'(T_0)]/\epsilon'(T_0)$,
(where $T_0(\omega)$ is a characteristic minimum) for vitreous SiO$_2$. It
can be seen that linear regimes in  $-\ln T$ for $T<T_0$ and $+\ln T$ for
$T>T_0$ are observed, and roughly with slopes $S_{-}=-2S$ and $S_{+}=S>0$, 
or in a -2:1 ratio. According to the 2LS TM, in fact, we have the expressions
\begin{eqnarray}
&&\left. \frac{\Delta\epsilon'}{\epsilon'} \right|_{2LS}=
\left. \frac{\Delta\epsilon'}{\epsilon'} \right|_{2RES}
+\left. \frac{\Delta\epsilon'}{\epsilon'} \right|_{2REL}, \label{dc2ls} \\
&&\left. \frac{\Delta\epsilon'}{\epsilon'} \right|_{2RES}=
\frac{2\bar{P}\overline{p_0^2}}{3\epsilon_0\epsilon_r}
\int_{z_{min}}^{z_{max}}\frac{dz}{z}
\sqrt{1-\left( \frac{\Delta_{0min}}{2k_BTz} \right)^2}\tanh z, 
\nonumber \\
&&\left. \frac{\Delta\epsilon'}{\epsilon'} \right|_{2REL}=
\frac{\bar{P}\overline{p_0^2}}{3\epsilon_0\epsilon_r}\times \nonumber \\
&&\times\int_{z_{min}}^{z_{max}}dz~\int_{\tau_{min}}^{\tau_{max}}
\frac{d\tau}{\tau}\sqrt{1-\frac{\tau_{min}}{\tau}}
\cosh^{-2}(z)\frac{1}{1+\omega^2\tau^2}, \nonumber
\end{eqnarray}
where we neglect, for low $\omega$, the frequency-dependence in the RES 
part, $z_{min,max}=\Delta_{0min,max}/2k_BT$ and where $\tau$ is the 
phenomenological 2LS relaxation time given by, with $E=2k_BTz$
\begin{equation}
\tau^{-1}=E\Delta_0^2/\gamma\tanh\left( \frac{E}{2k_BT}\right ).
\label{rt2ls}
\end{equation}

                                                                                
\begin{figure}[h]
\includegraphics[width=0.48\textwidth]{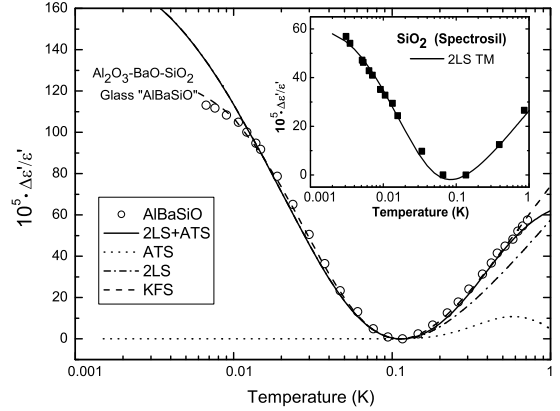} \vskip-5mm
\caption[2]{ Dielectric signature of pure $a$-SiO$_2$ (inset) and AlBaSiO
(main) glasses. SiO$_2$ data \cite{ref8}, fitted with Eq. (\ref{dc2ls}),
display a -2:1 2LS TM behavior. AlBaSiO data \cite{ref1} display rather
a -1:1 behavior, yet could be fitted with Eq. (\ref{dc2ls}) (dashed 
line) \cite{ref9} with a large $\Delta_{0min}=$12.2 mK 2LS tunneling 
parameter. We have fitted all data with a more realistic 
$\Delta_{0min}=$3.9 mK and best fit parameters from Table 1 using Eq.s 
(\ref{dc2ls}) and (\ref{dcats}) (driving frequency $\omega=$1 kHz). }
\label{fig1}
\end{figure}

Here, $\Delta_0$ is the tunneling parameter of a 2LS having energy gap
$E$, $\Delta_{0min}$ and $\Delta_{0max}$ are its phenomenological bounds, 
$\gamma$ is an elastic material parameter of the solid and 
$\tau_{min}^{-1}=E^3/\gamma\tanh\left( \frac{E}{2k_BT} \right)$,
$\tau_{max}^{-1}=E\Delta_{0min}^2/\gamma\tanh\left( \frac{E}{2k_BT}
\right)$. $\bar{P}$ is the probability per unit volume and energy that 
a 2LS occurs in the solid and $\overline{p_0^2}$ is the average square 
2LS electric dipole moment.

Indeed, from expressions (\ref{dc2ls}) we see: 1) The so-called resonant 
(RES) contribution has the leading behavior 
\begin{eqnarray}
\left. \frac{\Delta\epsilon'}{\epsilon'} \right|_{2RES}\simeq\cases{
-\frac{2}{3}\frac{\bar{P}\overline{p_0^2}}{\epsilon_0\epsilon_r}
\ln\left( \frac{2k_BT}{\Delta_{0max}} \right) & if 
$T<\frac{\Delta_{0max}}{2k_B}$,\cr
0 & if $T>\frac{\Delta_{0max}}{2k_B}$;\cr
}
\end{eqnarray} 
2) the relaxational (REL) contribution has, instead, the leading behavior
\begin{eqnarray}
\left. \frac{\Delta\epsilon'}{\epsilon'} \right|_{2REL}\simeq\cases{
0 & if $\omega\tau_{min}\gg1$ \cr
\frac{1}{3}\frac{\bar{P}\overline{p_0^2}}{\epsilon_0\epsilon_r}
\ln\left( \frac{2k_BT}{\Delta_{0min}} \right) & if $\omega\tau_{min}\ll 1$. 
\cr }
\end{eqnarray}
Thus, the sum of the two contributions has a V-shaped form, in a 
semilogarithmic plot, with the minimum occurring at a $T_0$ roughly
given by the condition $\omega\tau_{min}(k_BT)\simeq 1$, or
$k_BT_0(\omega)\simeq (\frac{1}{2}\gamma\omega)^{1/3}$. 
$\epsilon_0\epsilon_r$ is here the bulk of the solid's dielectric constant 
and we see that a -2:1 characteristic behavior is justified by the TM with 
the $T>T_0$ slope given by 
$S=\bar{P}\overline{p_0^2}/3\epsilon_0\epsilon_r$. 

This behavior is observed in pure $a$-SiO$_2$ \cite{ref8} (with the 
parameters of Table 1, $x=0$, from our own best fit to Eq. (\ref{dc2ls})), 
however in most multicomponent glasses it is rather a V-shaped curve with 
a (roughly) -1:1 slope ratio that is often observed. Fig. 1 (main) shows 
this phenomenon for the multisilicate glass Al$_2$O$_3$-BaO-SiO$_2$ (in 
short, AlBaSiO), which has been extensively investigated in recent times 
due to its unexpected magnetic field response \cite{ref1}-\cite{ref3}. 
Also, Fig. 2 shows the behavior of the dielectric constant vs $T$ for the 
glasses of composition (SiO$_2$)$_{1-x}$(K$_2$O)$_x$ containing a molar 
concentration $x$ of alkali oxide \cite{ref4}. It is seen that a 
$S_{-}/S_{+}$ slope ratio of roughly -1:1 is observed, with the slope 
definitely changing with $x$ (and faster for $T>T_0$). These data, thus 
far unexplained by the 2LS TM, call for an extension of the accepted TM 
and we show in this Letter that a simple explanation can be given in 
terms of the very same new (anomalous) tunneling systems (ATS) that have 
been advocated by one of us to explain the magnetic response of AlBaSiO 
and other multicomponent glasses \cite{ref3,ref5}. In view of the interest 
for these materials in low-T thermometry, and on fundamental grounds, such 
explanation appears overdue to us. Moreover,``additional'' TS (beside the 
standard 2LS) of the type here advocated were already called for in 
\cite{ref4} and earlier papers.  

In a rather general fashion, the TS can be thought of as arising from the 
shape of the theoretical energy-landscape $E(\{ {\bf r}_i \})$ of a glass 
as $T$ is lowered below the glass freezing transition $T_f$. Many local and 
global minima develop in $E(\{ {\bf r}_i \})$ as $T\to 0$, the 
lowest-energy minima of interest being made up of $n_w=2, 3, \dots$ local 
wells separated by shallow barriers. These local multiwelled potentials are 
our TS and it seems reasonable that the $n_w=2$ - welled potentials will be 
ubiquitous in this picture. These should be thought of as effective 
representations of local ``tremblements'' of the equilibrium positions 
$\{ {\bf r}_i^{(0)} \}$ of some of the glass ions' positions (unlike in the 
disordered crystal case, where the TS ought to be rather well-localized 
dynamical entities). Hence, just as the $n_w=2$ - welled case is possible, 
so ought to be the $n_w=3, 4, \dots$ - welled situations which would also be 
local rearrangements involving a few atoms/ions. So long as their energy 
parameters obey the usual uniform distribution advocated by the TM, however, 
most of these $n_w$-welled potentials should present the very same physics 
as the $n_w=2$ cases and thus in practice the $n_w$ distribution cannot be 
resolved experimentally in a pure glass.

                                                                                
\begin{figure}[h]
\includegraphics[width=0.48\textwidth]{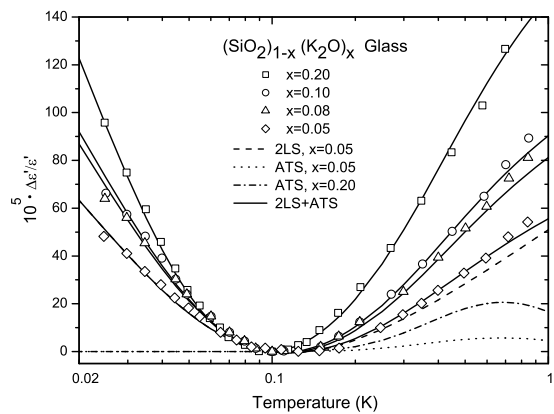} \vskip-5mm
\caption[2]{ Dielectric signature of mixed (SiO$_2$)$_{1-x}$(K$_2$O)$_x$
glasses as function of $T$ and $x$ \cite{ref4}. Fitting parameters from
Table 1 using Eq. (\ref{dc2ls}) and (\ref{dcats}) from our theory
(driving frequency $\omega=$10 kHz). }
\label{fig2}
\end{figure}

All change if the glass is made up by a mixture of network-forming (NF)
ions (like those of the SiO$_4$ or (AlO$_4$)$^{-}$ tetrahedral groups) 
{\em as well as} of network-modifying (NM) ions (like K$^{+}$ or Na$^{+}$, 
or Ba$^{2+}$, Pb$^{2+}$, ...) which, these last ones, could act as 
nucleating centres for a partial {\em devitrification} of the glass, as 
is know to occur in the multicomponent materials \cite{ref10,ref11}. 
Simulations and experiments in the multisilicates have shown that the
NM-species in part destroy the networking capacity of the NF-ions and form 
their own pockets and channels inside the NF-network \cite{ref12}. Hence, 
$n_w>2$ multiwelled systems inside these NM-pockets and -channels should 
follow some new energy-parameters' distribution forms when some degree of 
devitrification occurs, leading to entirely new physics.

One of the present Authors has proposed that precisely this situation 
occurs inside the magnetic-sensitive multicomponent glasses 
\cite{ref3,ref5}, and in this Letter we show how this theory explains the 
$B=0$ dielectric data of Fig.s 1-3 as well. Using the theory of 
\cite{ref3,ref5} we arrive at the expression, for the contribution to the 
dielectric anomaly from the advocated ATS:
\begin{eqnarray}
&&\left. \frac{\Delta\epsilon'}{\epsilon'} \right|_{ATS}=
\left. \frac{\Delta\epsilon'}{\epsilon'} \right|_{ARES}
+\left. \frac{\Delta\epsilon'}{\epsilon'} \right|_{AREL}, \label{dcats} \\
&&\left. \frac{\Delta\epsilon'}{\epsilon'} \right|_{ARES}=
\frac{\pi\tilde{P}^*\overline{p_1^2}}{3\epsilon_0\epsilon_rD_{min}}
\int_1^{\infty}\frac{dy}{y^2}\tanh\left( \frac{D_{min}}{2k_BT} y \right), 
\nonumber \\
&&\left. \frac{\Delta\epsilon'}{\epsilon'} \right|_{AREL}=
\frac{\pi\tilde{P}^*\overline{p_1^2}}{2\epsilon_0\epsilon_rD_{min}}
\left( \frac{D_{min}}{2k_BT} \right)\times 
\nonumber \\
&&\times\int_1^{\infty}\frac{dy}{y}\cosh^{-2}\left( 
\frac{D_{min}}{2k_BT} y \right)\frac{1}{1+\omega^2\tau_{Amax}^2}. 
\nonumber
\end{eqnarray}
Here we have again neglected, for low-$\omega$, the frequency-dependence 
in the RES part, we put $y=D/D_{min}$ and $\tau_{Amax}$ is the largest
phenomenological ATS relaxation time given by
\begin{equation}
\tau_{Amax}^{-1}=D^5/\Gamma\tanh\left( \frac{D}{2k_BT}\right ).
\label{rtats}
\end{equation}
Moreover $D$ is the lowest-lying energy gap of a multilevel ATS, $\Gamma$ 
is an appropriate elastic constant and $\tilde{P}^*$ is the (slightly 
renormalised) probability per unit volume that an ATS occurs in the NM
pockets and channels, with $\overline{p_1^2}$ the average square ATS dipole
moment. This description is linked to a distribution function 
\cite{ref3,ref5} for these ATS favoring near-degenerate energy gaps $D$ 
bound from below by $D_{min}$. In turn, this produces an overall density 
of states 
$g(E)=g_{2LS}+g_{ATS}(E)\simeq 2\bar{P}+(2\pi\tilde{P}^*/E)\theta(E-D_{min})$ 
that is roughly of the form advocated in \cite{ref4} and by some other 
earlier Authors (e.g. \cite{ref13}) to explain anomalies not accounted 
for by the standard 2LS TM.    
   
Manipulation of the expressions in (\ref{dcats}) shows that: 1) The RES 
contribution from the ATS has the leading behavior (for $T<D_{min}/2k_B$, 
$\epsilon'|_{ARES}$ is a constant)
\begin{eqnarray}
\left. \frac{\Delta\epsilon'}{\epsilon'} \right|_{ARES}\simeq\cases{
0 & if $T<\frac{D_{min}}{2k_B}$, \cr
\frac{\pi\tilde{P}^*\overline{p_1^2}}{6\epsilon_0\epsilon_rk_BT} 
\ln\left( \frac{2k_BT}{D_{min}} \right) & if 
$T>\frac{D_{min}}{2k_B}$; \cr
}
\end{eqnarray}
2) the REL contribution is, instead, characterised by the leading form
\begin{eqnarray}
\left. \frac{\Delta\epsilon'}{\epsilon'} \right|_{AREL}\simeq\cases{
0 & if $\omega\tau_{Amax}\gg1$ \cr
\frac{\pi\tilde{P}^*\overline{p_1^2}}{\epsilon_0\epsilon_rk_BT}
\ln\left( \frac{k_BT}{D_{min}} \right) & if $\omega\tau_{Amax}\ll 1$. 
\cr }
\end{eqnarray}
Thus, the V-shaped semilogarithmic curve is somewhat lost. However
adding the 2LS (Eq. (\ref{dc2ls})) and ATS (Eq. (\ref{dcats})) 
contributions together one does recover a V-shaped curve with a slope 
$S_{-}\simeq -2S$ basically unchanged for $T<T_0$ and an augmented slope 
$S_{+}=S+S_{ATS}$ for $T>T_0$ with 
$S_{ATS}=7\pi\tilde{P}^*\overline{p_1^2}/6\epsilon_0\epsilon_rk_BT$ that 
for $T<D_{min}/k_B$ may approach $2S$ and thus (qualitatively) a -1:1 slope 
ratio.


\begin{figure}[h]
\includegraphics[width=0.48\textwidth]{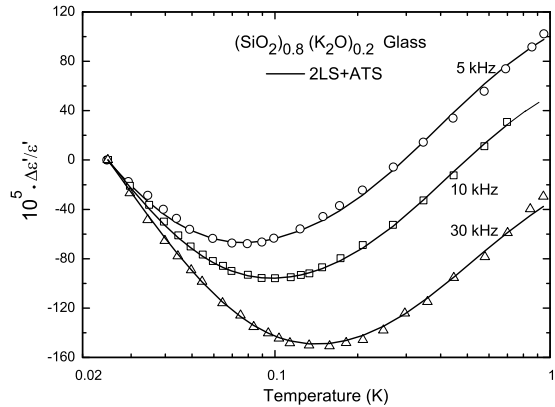} \vskip-5mm
\caption[2]{ Dielectric signature of mixed (SiO$_2$)$_{1-x}$(K$_2$O)$_x$
glasses as function of $T$ and $\omega$ for $x=$0.2 \cite{ref4}. Fitting 
parameters from Table 1 using Eq. (\ref{dc2ls}) and (\ref{dcats}) from our 
theory. }
\label{fig3}
\end{figure}

\begin{table}[h]
\begin{tabular}{|l|cccccc|}
\hline
glass & $x$ & $A_{2LS}$ & $\gamma$ & $A_{ATS}$ &
$D_{min}$ & $\Gamma$ \\
type & $$ & 10$^{-5}$ & 10$^{-8}$ sJ$^3$ & 10$^{-5}$ & K & 
10$^{-6}$ sK$^5$ \\
\hline \hline
SiO$_2$ & 0 & 47.2  & 5.30 & -  & -  & - \\
\hline
AlBaSiO & - & 116.2 & 13.40 & 264.7 & 0.65 & 69.73 \\
\hline
K-Si & 0.05 & 104.1 & 1.33 & 75.5 & 0.87 & 3.55 \\
\hline
K-Si & 0.08 & 146.5 & 1.23 & 130.0 & 0.87 & 3.97 \\
\hline
K-Si & 0.10 & 158.5 & 1.15 & 160.0 & 0.87 & 5.08 \\
\hline
K-Si & 0.20 & 239.5 & 0.82 & 281.9 & 0.87 & 6.44 \\
\hline
\end{tabular}
\caption[1]{ Extracted parameters for the glasses; K-Si stands for the 
(SiO$_2$)$_{1-x}$(K$_2$O)$_x$ glasses. In all of the best fits we have 
employed the values $\Delta_{0min}=$3.9 mK and $\Delta_{0max}=$10 K 
extracted from fitting the pure SiO$_2$ data of Fig. 1 (inset).}
\label{tabl1}
\end{table}

We have fitted expressions (\ref{dc2ls}) and (\ref{dcats}) to the data
for AlBaSiO in Fig. 1 (main) and to the $x$-dependent data for 
(SiO$_2$)$_{1-x}$(K$_2$O)$_x$ in Fig.s 2 and 3, obtaining in all cases 
very good agreement between theory and experiments \cite{ref14}. Fig. 3 
shows the fit of our theory to the frequency-dependent data for $x=0.2$. 
In all these best fits we have kept the value of $\Delta_{0min}=3.9$ mK 
fixed, as obtained from our pure SiO$_2$ fit, and the value of $D_{min}$ 
also independent of $x$ and $\omega$. The idea is that these parameters are 
rather local ones and should not be influenced by NF/NM dilution. Table 1 
reports all the (2LS and ATS) parameters used for our best fits and for 
AlBaSiO the parameters differ somewhat from the values 
$A_{ATS}=1.42\times$10$^{-3}$ and $D_{min}=0.03$ K used in \cite{ref5}. 
This shows that a more microscopic version of this model is called for.   
Fig. 4 shows the dependence of the prefactors with $x$. It can be seen that, 
as expected, the ATS prefactor 
$A_{ATS}=\pi\tilde{P}^*\overline{p_1^2}/\epsilon_0\epsilon_rD_{min}$ 
scales linearly with $x$, an excellent confirmation that the 
``additional'' TS of \cite{ref4,ref13} are indeed our ATS forming near 
the microcrystallites within the NM-pockets and channels. It can be seen, 
instead, that the 2LS prefactor 
$A_{2LS}=\bar{P}\overline{p_0^2}/\epsilon_0\epsilon_r$ of our fits also 
increases, though less rapidly, with $x$ (a decrease like $1-x$ would be 
expected). We propose (adopting a NF/NM percolation picture) that new, 
``induced'' 2LS form with alkali dilution near the NF/NM surface of the 
NF percolating clusters as $x$ is increased from 0. This leads to the 
expression $A_{bulk}(1-x)+A_{surf}P(x)x^f$ for the 2LS prefactor, with 
$A_{bulk}$, $A_{surf}$ and $f$ fitting parameters and $P(x)$ the 
percolation probability function ($P(x)\simeq 1$ for small $x$). Our 
best fit leads to $f=0.81$, in rather good agreement with the euristic 
expression $f=1-(D-D_s)\nu$ ($D$ is the fractal dimension of the 
percolating cluster, $D_s<D$ of its ``elastic'' surface (not necessarily
the hull), $\nu$ the connectedness length's exponent) one would infer 
from elementary fractal or percolation theory.


\begin{figure}[h]
\includegraphics[width=0.48\textwidth]{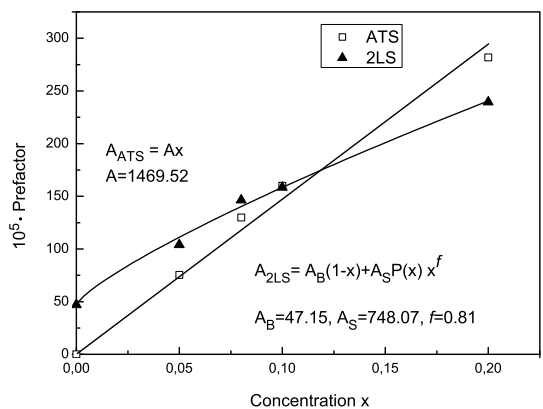} \vskip-5mm
\caption[2]{ The 2LS and ATS prefactor parameters ($\times 10^5$) for all 
glasses (from Table 1) as a function of $x$. Our data fit well with our 
theoretical expectations (full lines). } 
\label{fig4}
\end{figure}

Our theory, with parameters from Table 1, can also account for the $T-$ 
and $x-$dependence of the heat capacity $C_p(T,x)$ (data from \cite{ref4};
for AlBaSiO see also \cite{ref3}).

In summary, we have shown that there is direct evidence in zero magnetic
field already for the multiwelled ATS advocated to explain the magnetic
field effect in the multicomponent glasses. The relevance of multiwelled 
tunneling systems in the multicomponent glasses is a new and unexpected 
finding in this field of research. Our work predicts that the magnetic 
response of the alkali silicate glasses should be important and scale like 
the molar alkali concentration $x$. At the same time the -1:1 slope ratio 
problem of the standard TM has been given a simple explanation in terms
of our two-component tunneling model. For AlBaSiO, using the value of 
$\tilde{P}^*$ extracted from $C_p$ \cite{ref3} and that of $A_{ATS}$ given
in Table 1, we estimate the value of the electic dipole of the effective
anomalous tunneling entity: 
$p_{eff}\equiv\sqrt{\overline{p_1^2}}\simeq$0.4 D. This small value 
confirms \cite{ref3,ref5} that this physics must come from the coherent 
tunneling of a small ionic cluster (the very same origin of the large 
value of $D_{min}$). Details of our analysis will be published elsewhere.

The present work suggests how new experiments on mixed glasses could help 
in deciding whether our two-component tunneling model \cite{ref3,ref5} or 
the nuclear electric quadrupole-moment approach \cite{ref15} is the 
correct explanation for the magnetic-field effects in the cold glasses. 
Amorphous solids of composition (SiO$_2$)$_{1-x}$(MO)$_x$ should be 
investigated for different concentrations $x$ of NM-species at ultralow
temperatures and in the presence of a magnetic field. If M is a divalent 
metal (ideally Ca, but also Mg, Sr and Ba) with no or a tiny percentage
of stable isotopes possessing a nonzero nuclear quadrupole moment, then a 
magnetic response scaling like $x$ would single out the present approach. 
The system (SiO$_2$)$_{1-x}$(CaO)$_x$ (with a small percent of PbO favoring 
nanocrystal formation) appears to be the ideal candidate. We remark, 
incidentally, that in \cite{ref2} the amorphous system SiO$_{2+x}$C$_y$H$_z$, 
devoid of nuclear quadrupole moments, displayed magnetic effects 
quantitatively similar to those found in glasses with nuclei carrying such 
moments \cite{ref1} at ultralow temperatures.


\begin{thebibliography}{99}


\bibitem{ref1} P. Strehlow, C. Enss and S. Hunklinger, Phys. Rev. Lett.
{\bf 80}, 5361 (1998); P. Strehlow, M. Wohlfahrt, A.G.M. Jansen,
R. Haueisen, G. Weiss, C. Enss and S. Hunklinger, Phys. Rev. Lett.
{\bf 84}, 1938 (2000) 

\bibitem{ref2} J. Le Cochec, F. Ladieu and P. Pari, Phys. Rev. B {\bf 66}, 
064203 (2002)

\bibitem{ref3} G. Jug, Phil. Mag. {\bf 84}, 3599 (2004)

\bibitem{ref4} W.M. MacDonald, A.C. Anderson and J. Schroeder, 
Phys. Rev. B {\bf 31}, 1090 (1985)

\bibitem{ref5} G. Jug, Phys. Rev. B {\bf 79}, 180201(R) (2009)

\bibitem{ref6} For recent reviews see: {P. Esquinazi (ed.) 
{\em Tunneling Systems in Amorphous and Crystalline Solids} (Springer, 
Berlin, 1998)}; C. Enss, Physica, {\bf 316B-317B}, 12 (2002); C. Enss, 
Adv. Sol. State Phys. {\bf 42}, 335 (2002) 

\bibitem{ref7} See, e.g., M.H. Cohen and G.S. Grest, Phys. Rev. Lett. 
{\bf 45}, 1271 (1980)

\bibitem{ref8} S.A.J. Wiegers, R. Jochemsen, C.C. Kranenburg and G. 
Frossati, Rev. Sci. Instrum. {\bf 58}, 2274 (1987)

\bibitem{ref9} S. Kettemann, P. Fulde and P. Strehlow, Phys. Rev. Lett.
{\bf 83}, 4325 (1999) (KFS in short)

\bibitem{ref10} C.-Y. Fang, H. Yinnon and D.R. Uhlmann, J. Non-Cryst.
Solids {\bf 57}, 465 (1983) 

\bibitem{ref11} N. Pellegri, E.J.C. Dawnay and E.M. Yeatman, J. Sol-Gel
Sci. and Tech. {\bf 13}, 783 (1998); nanocrystals are ubiquitous in the
multisilicate glasses, see e.g.: G. De {\it et al.}, J. Non-Cryst. Sol.
{\bf 194}, 225 (1996); X.L. Duan {\it et al.}, J. Cryst. Growth, {\bf 252},
311 (2003)

\bibitem{ref12} A. Meyer, J. Horbach, W. Kob, F. Kargl and H. Schober,
Phys. Rev. Lett. {\bf 93}, 027801 (2004); J. Reinisch and A. Heuer,
J. Phys. Chem. B {\bf 110}, 19044 (2006) 

\bibitem{ref13} C.C. Yu and A.J. Leggett, Comments Cond. Mat. Phys., 
{\bf 14}, 231 (1988); J.L. Black, Phys. Rev. B {\bf 17}, 2740 (1978)

\bibitem{ref14} TS-TS interactions can be neglected above $T\sim$ 100 mK.

\bibitem{ref15} A. W\"urger, A. Fleischmann and C. Enss, Phys. Rev. Lett.
{\bf 89}, 237601 (2002)

\end{thebibliography}
\end{document}